# Hydrogen Bond Symmetrization in Glycinium Oxalate under Pressure[§]

Himal Bhatt,[1] Chitra Murli,[1] A. K. Mishra,[1] Ashok K. Verma,[1] Nandini Garg,[1] M.N. Deo,[1] R. Chitra,[2] and Surinder M. Sharma[1,*]

[a]High Pressure and Synchrotron Radiation Physics Division, [b]Solid State Physics Division, Bhabha Atomic Research Centre, Trombay, Mumbai 400085, India

We report here the evidences of hydrogen bond symmetrization in the simplest amino acid- carboxylic acid complex, glycinium oxalate, at moderate pressures of 8 GPa using in-situ infrared and Raman spectroscopic investigations combined with first-principles simulations. The protonation of the semioxalate units through dynamic proton movement results in infinite oxalate chains. At pressures above 12 GPa, the glycine units systematically reorient with pressure to form hydrogen bonded supramolecular assemblies held together by these chains.

Hydrogenous materials such as water, organic acids and minerals at extreme thermodynamic conditions are of immense importance to the understanding of bio-geological processes in earth's interior as well as other celestial bodies. The search for primitive terrestrial life is based on the assembly of organic matter in extreme conditions, viz. composites of carbon dioxide, methane, water, ammonia and amino acids [1-2]. Recently, traces of glycine, the simplest amino acid, and its complexes have been found in hot molecular cores of star forming regions, Mars, meteorites and interstellar dust [1, 3].

Hydrogen bonds play a decisive role in the structural stabilization of these materials. For example, at high pressures, the structure of non-molecular phase of ice (~100 GPa) is stabilized by a symmetric hydrogen bond formed through "translational proton tunnelling" [4-5]. In the symmetrization limit ($d_{O---O} < 2.4$ Å) of the O-H---O hydrogen bonds, many exotic phenomena of fundamental interest occur, such as realization of a unimodal symmetric well at bond centre [4], equal proton sharing leading to polymerization [6] and structural distortion affecting electronic processes like spin cross-over [7], and ionic configuration through proton migration [8].

The study of hydrogen bonds near symmetrization limit is of importance in the larger context of understanding proton dynamics in the complex bio-geological processes in nature where inter-molecular interactions are governed by diverse chemical environments [9]. Though, there have been efforts to study the dynamical behavior of O-H---O bond configuration in carboxylic acid complexes [6, 8], hydrogen halides [10], salt waters [11] and oxyhydroxide minerals [7] under pressure, only in water ice, strong evidences of hydrogen bond symmetrization exist. Alternatively, the strong hydrogen bonded systems such as amino acid- oxalic acid complexes provide unique opportunity to study proton dynamics near the symmetrization limit due to their extreme sensitivity to hydrogen bond tunability under pressure [12]. The underlying mechanism of proton dynamics in these complexes can be used to understand proton transfer pathways in protein environments and molecular theories of ionic systems at extreme conditions [13]. The hydrogen bonded networks in these materials are also ideal to look for novel structures like dynamic polymers [14] through hydrogen bond assisted supramolecular assembly [15-17], a path adopted in nature for bio-material synthesis [18].

Glycinium oxalate (GO), the simplest amino acid-carboxylic acid complex crystallizes in monoclinic structure with space group $P2_1/c$ and four formula units per unit cell (Z=4) [19]. It possesses a nearly linear and strong O3-H7---O6 hydrogen bond ($d_{O---O}$ ~2.54 Å, $d_{H---O}$ ~1.6 Å and $\angle OHO = 177°$) between semioxalate molecules in a columnar arrangement along the b-axis (Fig. 1bottom). The strengthening of this hydrogen bond, under pressure, to the symmetrization limit may lead to interesting proton dynamics in this compound. These semioxalate columns hold the head to tail linked glycine sheets in the ac-plane via other hydrogen bonds to form the three dimensional network (Fig. 1top).

In this letter, we report hydrogen bond symmetrization in GO through proton sharing in the O3-H7---O6 hydrogen bond at moderate pressures using combined high pressure Raman and infrared (IR) spectroscopic measurements up to ~20 GPa and first-principles DFT based molecular dynamics (MD) simulations [S2-S5 in ref. 20]. The strengthening of the O3-H7---O6 hydrogen bond under pressure to the symmetrization limit is evident from the bond parameters obtained from MD calculations (Fig. 2). A total of 12000 equilibrated configurations were used at each pressure to generate the dynamical picture. It shows that the spread of the proton distribution reduces with pressure and shifts towards the bond mid-point (i.e., $\delta = 0$). As the $d_{O3---O6}$ continuously reduces to ~2.4 Å at pressures close to 10 GPa, a systematic increase in $d_{O3-H7}$ and decrease in $d_{H7---O6}$ is noted (Fig. 2d), while $\angle OHO$ (178°) remains close to ~180°. We can further see from Fig. 2d that the probability of crossing the mid-point through proton hopping, increases with pressure and is already significant at ~ 10 GPa.

During such proton movement towards the mid-point of the hydrogen bond, the system becomes highly anharmonic. This results in remarkable dampening of νOH (stretching) mode and cascading interactions between vibrational energy levels [21-23]. In GO, as the νOH IR mode (~ 2360 cm$^{-1}$) [24] lies in the second order diamond absorption region, a first hand information on its high pressure behaviour was obtained by phonon calculations using quantum espresso code. The pressure induced softening of this mode, as obtained from Lorentz oscillator fit at ambient pressure and phonon calculations up to 2.5 GPa (Fig. 3c), could be approximated by the formula $\nu = [A(P_c-P)]^m$ [6], which gives A~1.72 x10$^5$, m ~ 0.53 and $P_c$ ~ 8.15 GPa as the pressure corresponding to νOH mode instability (Fig. 3b inset). Thus, large red shift in νOH (> 250 cm$^{-1}$ upto 2.5 GPa) and substantial reduction in the O3---O6 distance ($d_{O3-O6}$ < 2.5 Å) (Fig. 2d) implies the approaching of symmetrization limit near 2 GPa.

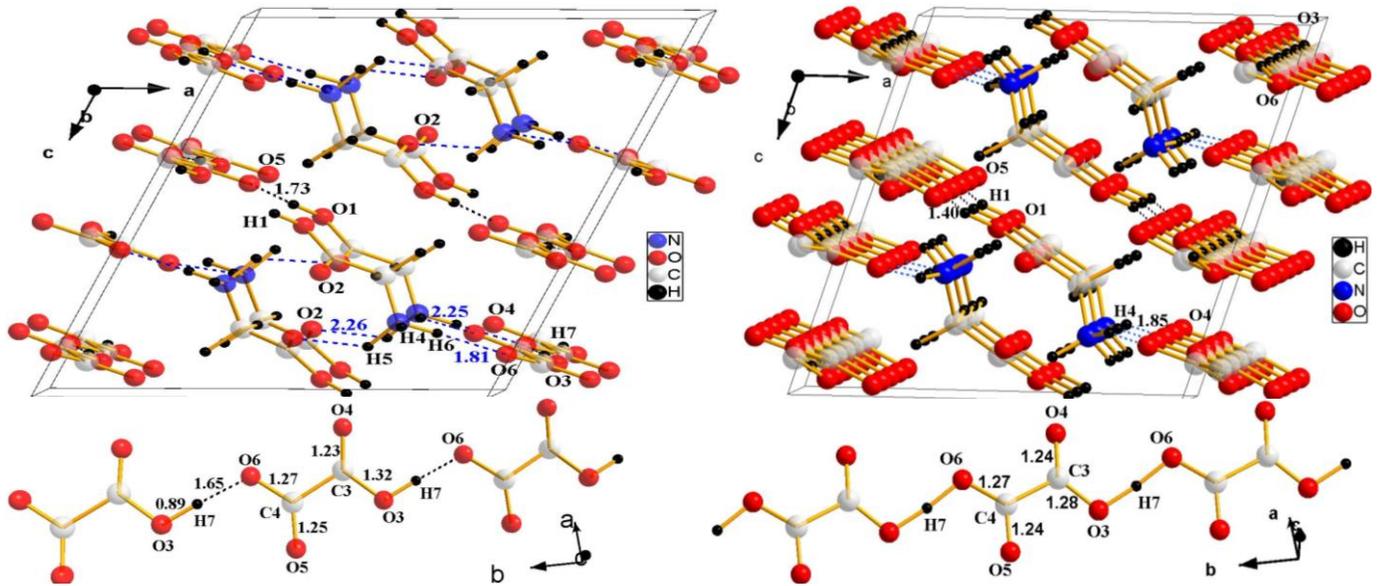

**FIG. 1.** (color online) Top left: Unit cell of GO with dashed lines for strong hydrogen bonds (blue: NH---O, black: OH---O, numbers are in Å units). Two glycine molecules are linked from head to tail via N1-H5---O2 in the *ac* plane forming bilayers. Two semioxalate columns along *b*-axis, linked through inversion symmetry connect two glycine columns also along *b*-axis through strong O1-H1---O5, N1-H6---O6 (H---O=1.81 Å) and N1-H4---O4 (H---O=2.25 Å) hydrogen bonds. Bottom left: One semioxalate column along b-axis bridged through strong O3-H7---O6 hydrogen bond. Top right: Structure of GO at 20 GPa, showing supramolecular assembly through O1H1-O5/N1H4-O4 hydrogen bonds, rest of the linking NH---O hydrogen bonds get weakened with pressure. Bottom right: Snapshot of simulated structure along *b*-axis (from MD) at 20 GPa showing symmetrization of O3-H7-O6 hydrogen bond, with average C-O/C=O distances.

Above 2 GPa, increase in anharmonicity is indicated by a relative increase in the background of the mid-IR band profile [25-26], as shown in Fig 3a. Across 8 GPa, the width of this band, riding over the background of various other fundamentals, significantly increases and its relative intensity grows. The centroid of this band shifts to lower frequencies, at an increased rate above 8 GPa, and is found at ~ 950 cm$^{-1}$ at 18 GPa. Such broad envelopes, resembling a continuum arise due to large νOH dampening and are used as prime evidence of strengthening of hydrogen bond in the low barrier hydrogen bonded systems close to symmetrization [21-22, 25-26], based on experiments and anharmonic calculations on various systems [22, 27].

The anomaly in the pressure variation of δOH (in-plane bend) (1232 cm$^{-1}$) oxalate mode near 5 GPa (Fig. 3b) and narrow transmission dips (! in Fig. 3a) resembling Evans hole, are spectral features arising due to the OH instability and mode couplings [21-22, 26, 28]. The significant stiffening of the γOH (out of plane bend) (~1018 cm$^{-1}$) and that of δOH modes (Fig. 3b) also support strengthening of this hydrogen bond under pressure [26, 29-30]. All these observations are reproducible and reversible on pressure

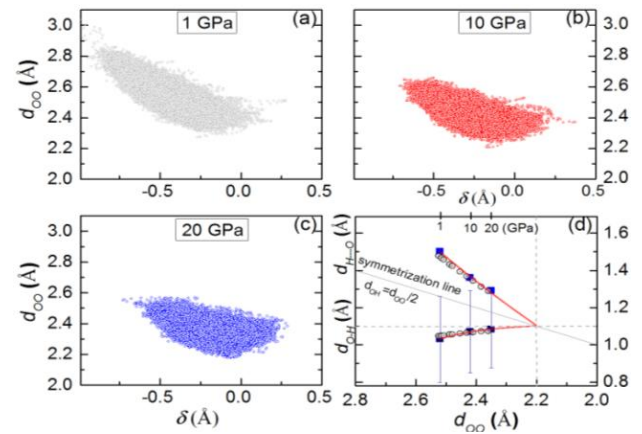

**FIG. 2.** (color online) (a-c) Evolution of dynamic spread in proton distribution function with pressure towards the bond midpoint ($\delta = d_{O-H} - d_{H---O}$) (d) Bond parameters of O3-H7---O6 (open circles-0K DFT; filled squares-300K first-principles MD); vertical bars represent spread in the $d_{O-H}$ values, this is linked to the probability of crossing the potential barrier for the classical proton. The fitted curves meet near 60 GPa if the initial phase persists [2, 4]. Proton quantum tunnelling may however reduce this pressure.

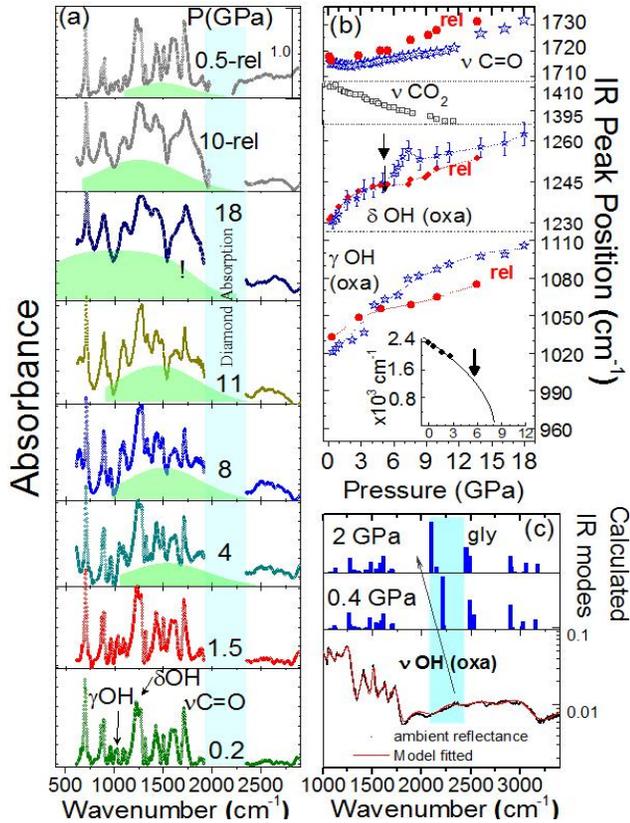
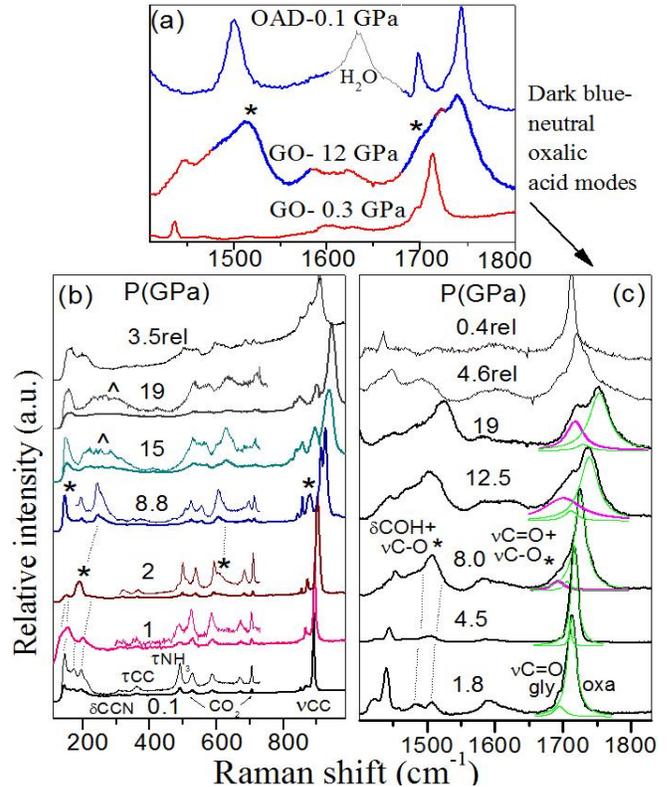

**FIG. 3.** (color online) (a) IR spectra (600 – 2900 cm$^{-1}$) of GO at various pressures. The highlighted hump due to νOH is guide to the eye (b) Variation of selected IR modes with pressure. Here red solid circles are release (rel) pressure data. Inset: a power law fit to the νOH mode up to 2.5 GPa and arrow indicates the position of intersection with δOH mode. (c) Ambient IR reflectance spectrum and calculated IR active modes in GO with pressure.

release.

However, the spectral signatures of the high pressure phase above 8 GPa are well preserved up to ~ 3 GPa on release (Fig. 3b), showing hysteresis behavior. Such a behavior has been attributed to the enhancement of covalent like O3-H7---O6 interactions between the semi-oxalate units in the high pressure phase [6]. This covalent nature would result from the proton sharing between the closely overlapped potential minima in which the proton can easily hop between the two sites as shown by dynamical simulations (Fig. 2). A nearly linear geometry of the hydrogen bond O3-H7---O6 at ambient as well as at higher pressures, i.e. ∠OHO ~180°, favors this mechanism. Our high pressure Raman spectra, which are discussed in the following section, indeed reveal proton sharing between the semioxalate units, which mimics the proton being at the mid-point, implying a symmetric state.

The proton sharing (hopping) along the O3-H7-O6 bond would result in the protonation of semioxalate unit (COO$^-$) and thus transforming it to neutral oxalic acid.

**FIG.4.** (color online) (a) comparison of oxalate Raman modes of the high pressure symmetrized phase of GO with neutral oxalic acid dihydrate (OAD); * denote neutral oxalate modes; High pressure Raman spectra of GO in (a) 100–750 cm$^{-1}$ and (b) 1400–1850 cm$^{-1}$ region.

Above 8 GPa, as shown in Fig. 4a, the spectral signatures of semioxalate in GO are replaced by the new characteristic modes of neutral oxalic acid around 1500 cm$^{-1}$ (δCOH+νC-O) and 1700 cm$^{-1}$ (νC=O+νC-O mode). This is also confirmed by comparing it with the ambient spectrum of a well studied neutral oxalic acid compound OAD (see Fig. 4a) and the new modes appeared are in close agreement with the neutral oxalic acid modes [30-31].

In addition, large blue shift (~ 41 cm$^{-1}$ up to 8 GPa) of νC=O oxalate mode (~1709 cm$^{-1}$) (Fig. 4c), discontinuous reduction (~7 cm$^{-1}$) in δOH IR mode across 8 GPa (Fig. 3b) and new features in δCO$_2$ region (near 600 cm$^{-1}$) (Fig. 4b) are spectral features arising due to the transformation of singly ionized to neutral oxalic acid under pressure [29-32]. Also, the prominent features of neutral oxalate structure like δC-COOH mode (~880 cm$^{-1}$) show abrupt increase in its relative intensity across 8 GPa. In support of the IR results, we also note distinctive features linked with strengthening of short hydrogen bonds in semioxalate complexes [20, 29], viz. large stiffening (2.8 cm$^{-1}$/GPa) of νC-C oxalate Raman mode (893 cm$^{-1}$) and appearance of a mode at ~200 cm$^{-1}$, close to that of νOHO hydrogen bond stretching vibrations (Fig. 4b) [30].

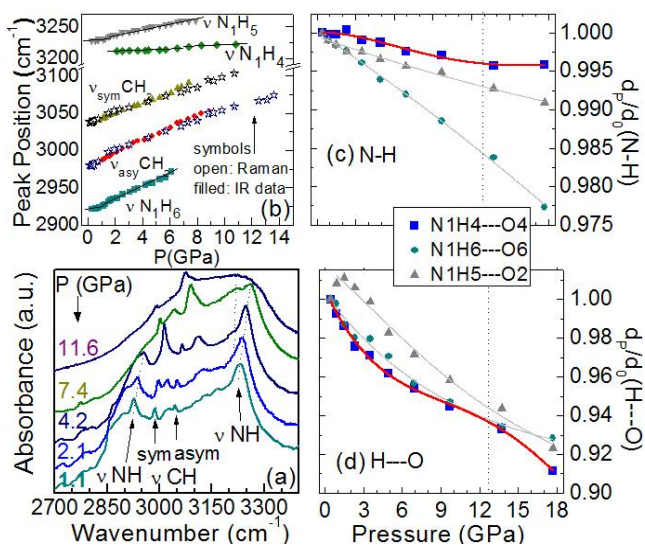

**FIG.5.** (color online) (a) High pressure IR spectra in CH/NH stretching region (2700–3400cm$^{-1}$) with various fundamental and combination/overtone modes. Dotted lines in νNH modes are guide to eye (b) Frequency vs. pressure plots for νCH and νNH modes (c) Variation of normalized N-H and (d) H---O distances (calculated) for various N-H---O hydrogen bonds with pressure

These spectral changes also imply transformation of C4-O5 group to C4=O5 on the acceptor site (see Fig.1 bottom) upon protonation. This transition is well supported by theory, as the calculated C-O (C4-O6; C3-O3) and C=O (C4=O5; C3=O4) bond parameters of oxalate units also approach values corresponding to neutral oxalic acid (Fig. 1). Also note that the resulting anti-planar neutral oxalate motif is an energetically favorable conformation [33]. Thus, in the high pressure phase, the neutral oxalate motifs formed due to symmetrized hydrogen bond, result in infinite chains along the *b*-axis.

These chains hold the glycine sheets in the *ac*-plane through strong O1-H1---O5 and N-H---O hydrogen bonds (Fig.1top left). The enhanced π bond character (C4=O5) of C4-O5 bonds of oxalates subsequent to protonation would result in strengthening of O1-H1---O5=C4 hydrogen bond [S6, ref. 20] between glycine and oxalic acid at further higher pressures.

Glycine molecule, towards the other end, is connected to the C3=O4 unit of carboxyl group of another oxalate chain through N1-H4---O4 hydrogen bonds. Interestingly, the νN1-H4 IR mode remains nearly invariant under pressure (Fig. 5b). The corresponding covalent bond length (N1-H4) shows relatively very small decrease with pressure (Fig. 5c). Thus, the reduction in $d_{H4---O4}$ at higher pressures (Fig. 5d) suggests that this N1-H4---O4=C3 hydrogen bond may be a blue shifting hydrogen bond, which becomes stronger at higher pressures as compared to the other N-H---O hydrogen bonds. Above 12 GPa, decrease in the relative intensity of νC=O oxalate mode (corresponding to C3=O4) (Fig. 4c) indicates reduction in the π bond character of C3=O4 and the formation of N1-H4---O4-C3 chain. In addition, the formation of a planar conformation of glycine (Fig. 1top right) suggested by the structural simulations and spectral changes further support the above mentioned propositions [S7, ref. 20].

At pressure above 12 GPa, the evolution of broad envelopes in the lattice (marked as ^ in Fig. 4a) and νN-H regions in both IR (Fig. 5a) and Raman spectra [S7, 20] indeed suggest emergence of a supramolecular phase. Thus, the stronger O1-H1---O5 and N1-H4---O4 interactions at high pressures result in glycine hammocks (in *ac*-plane) tied to the infinite oxalate poles (*b*-axis) as shown in Fig. 1 (top right).

To summarize, we have reported hydrogen bond symmetrization in glycinium oxalate, the simplest amino acid-carboxylic acid molecular system, through computational molecular dynamics of O-H---O bond, infrared studies of O-H instability and Raman studies of proton sharing at pressures as low as 8 GPa. We have also demonstrated the role of proton hopping in an O-H---O hydrogen bond close to the symmetrization limit. During large system anharmonicity, an incipient proton sharing results from the evolution of the average proton distribution towards the bond mid-point. In such situations, the bond parameters adjacent to O-H---O units and their corresponding spectroscopic parameters approach values in accordance with the proton localized unimodal symmetric state. So far, there have been very few systems reported in this context except for extensive investigations on the simplest O-H---O system ice. The biological and geological systems in nature are quite complex, with diverse chemical environments governing their inter-molecular interactions. Hence, this work would be very useful for understanding the physics of hydrogen bond symmetrization and proton dynamics in a broader context of organic systems. Our reports of supramolecular structures formed through tuning of hydrogen bonds under pressure are also of relevance to organic linkages at extreme conditions and dynamic polymers with non covalent interactions.

# Hydrogen Bond Symmetrization in Glycinium Oxalate under Pressure

Himal Bhatt,[1] Chitra Murli,[1] A. K. Mishra,[1] Ashok K. Verma,[1] Nandini Garg,[1] M.N. Deo,[1] R. Chitra,[2] and Surinder M. Sharma[1,*]

**Supplementary information**

## S1. Synthesis:

Colorless three-dimensional transparent single crystals of glycinium oxalate (GO) were grown by mixing α-glycine and oxalic acid dihydrate in a 1:1 stoichiometric ratio in water. Two clear crystals of approximately $3\times1\times0.5$ mm$^3$ were selected and used for all the measurements.

## S2. Spectroscopic investigations:

**Vibrational Mode Assignments:**

Mode assignments of glycinium oxalate (GO) have been carried out based on earlier IR, Raman and theoretical studies of GO [34], Bis(glycinium)oxalate (BGO) [35], tri glycine sulphate (TGS) [36], α-glycine [37] and oxalic acid dihydrate [30-31, 38-40]. The assignments were also checked using reported values under ambient conditions for various other glycine and oxalate complexes [12, 26, 41-46] and based on the high pressure behavior of IR and Raman active modes in GO [this study], BGO [35] and oxalic acid dihydrate (OAD) [39]. The mode assignments for oxalic acid have further been verified based on the extensive works by Novak and coworkers on various oxalate and semioxalate compounds possessing hydrogen bonds of varying strengths at ambient conditions [25, 38, 47-49]. For OH and NH stretching IR modes, the correlation curves reported for stretching frequencies as a function of hydrogen bond strength, i.e. bond parameters have been used [24]. The modes assignments were finally verified using our first-principles calculations which are also in agreement with the reported values.

**Infrared spectroscopy**

For infrared absorption studies, an indigenously developed clamp type Diamond Anvil Cell (DAC) was mounted on the sample stage of Hyperion-2000 IR microscope, coupled to the Bruker Vertex 80V FTIR equipped with Globar source and KBr beamsplitter. A liquid nitrogen cooled MCT detector was used for the complete mid infrared range. Polycrystalline sample in CsI matrix along with a ruby ball was loaded in a 150 μm hole of a tungsten gasket pre-indented to a thickness of ~60 μm. Ruby balls were kept both near the centre and at the periphery of the diamond cell. Pressure calibration was done using Ruby fluorescence lines [50]. Empty cell spectra and that of CsI loaded in the DAC were used to deduce the sample transmittance. For clear identification of all the spectral features in the 600 – 4000 cm$^{-1}$ spectral range, the spectra were recorded at 2 cm$^{-1}$ resolution and some of the repeat measurements were carried out at 4 cm$^{-1}$ resolution. The ambient pressure spectrum of GO was also recorded using Bruker IFS125 HR FT-spectrometer and all the peak positions recorded from the two spectrometers show excellent agreement.

**Raman spectroscopy**

Raman studies were carried out using a confocal micro Raman set up having a HR 460 Jobin Yvon single stage spectrograph equipped with LN$_2$ cooled Spectrum-One CCD detector. The spectra at each pressure were calibrated using standard neon lines [50-51]. GO crystal (~ 50 μm) along with a couple of ~ 10 μm ruby balls were loaded in the pre-indented tungsten gasket of thickness ~70 μm with a hole of diameter ~100 μm in a modified Mao-Bell type of Diamond anvil cell (DAC). In the repeat measurements, ruby balls were kept at the centre as well as the peripheries. The ambient pressure spectrum of GO was also recorded using a Bruker MultiRAM FT-Raman spectrometer and all the peak positions in ambient pressure spectra recorded from the two spectrometers were in excellent agreement.

**S3. Theoretical simulations:**

All the simulations were performed using the Vienna Ab Initio Simulation Package (VASP) [52-55]. The interactions between core and valence electrons were treated with projector plane wave (PAW) method with $s^1$ (H), $s^2p^2$ (C), $s^2p^3$ (N) and $s^2p^4$ (O) valence configurations. The exchange-correlation energy was treated with PBE version of GGA and an energy cut-off of 600 eV was used for plane wave basis construction [56].

In the case of structural optimization, the Brillouin zone was sampled by 4×6×4 Monkhorst-Pack mesh whereas Γ-point was used for MD simulations [57]. The MD simulations were performed for a 1×2×1 (144 atoms) supercell in the canonical (NVT) ensemble and the temperature was controlled using the Nosè thermostat. The MD time-step was taken equal to 1 femto-second. All simulations run for 15 ps. Initial 3 ps data were not used in the analysis.

The vibrational properties were studied using density functional perturbation theory as implemented in the Quantum-Espresso computer code [58]. For these calculations, we have used the local density approximation for exchange-correlation [59] as the code does not allow the Raman and infrared intensities calculation for GGA exchange-correlation. An energy cutoff of 160 Ry was used for plane wave expansion. The Brillouin zone was sampled using same Monkhorst-Pack k-point grid as in structural relaxations. The crystal structures were fully optimized before these calculations.

## S4. High pressure x-ray diffraction studies

Angular Dispersive X-ray Diffraction (ADXRD) studies were carried out at the 5.2R (XRD1) beamline of Elettra synchrotron source with monochromatized x-rays ($\lambda \sim 0.6888$Å). For these experiments, the polycrystalline sample along with few particles of Cu, was loaded in a ~120 μm diameter hole drilled in a pre-indented tungsten gasket of a Mao-Bell type of DAC and no pressure transmitting medium was used. Pressure was calculated using Birch-Murnaghan equation of state of Cu [60]. The diffraction patterns were recorded using a MAR345 imaging plate detector kept at a distance of ~ 20 cm from the sample. The diffraction profiles were obtained by the radial integration of the two dimensional diffraction rings using the FIT2D software [61].

## S5. Structural investigations

The observed x-ray diffraction (XRD) patterns (Fig. SM1) match with the Bragg reflections corresponding to the monoclinic structure with space group $P2_1/c$ and four formula units per unit cell (Z=4). The ambient pressure lattice parameters obtained using Rietveld refinement [62] are: a = 10.614 (4) Å, b = 5.649 (3) Å, c = 12.096 (5) Å, α = γ = 90° and β = 113.769 (3), which are in close agreement with the earlier published values [19]. At high pressure, as the intensity of the XRD peaks of GO is poor, experimental lattice parameters were determined only up to 1.8 GPa using Le'Bail refinement incorporated into GSAS. The unit cell volumes at different pressures, determined from x-ray diffraction experiments and theoretical calculations agree reasonably well, except for a small deviation at very low pressures. The linear compressibilities ($-(1/l) \times (dl/dp)$) are determined to be $5.5 \times 10^{-2}$ /GPa, $2.6 \times 10^{-2}$ /GPa and $0.92 \times 10^{-2}$ /GPa using synchrotron x-ray diffraction data up to 1.8 GPa ($5.3 \times 10^{-2}$ /GPa, $2.4 \times 10^{-2}$ /GPa and $0.9 \times 10^{-2}$ /GPa, from theory) in the c, a and b directions respectively, which show anisotropic contraction with b-direction showing least compression (Fig. SM2). This is in good qualitative agreement with the theoretically calculated values. The full width at half maximum (FWHM) of x-ray diffraction peaks slightly increases with pressure. Ambient structure could be retrieved on release of pressure.

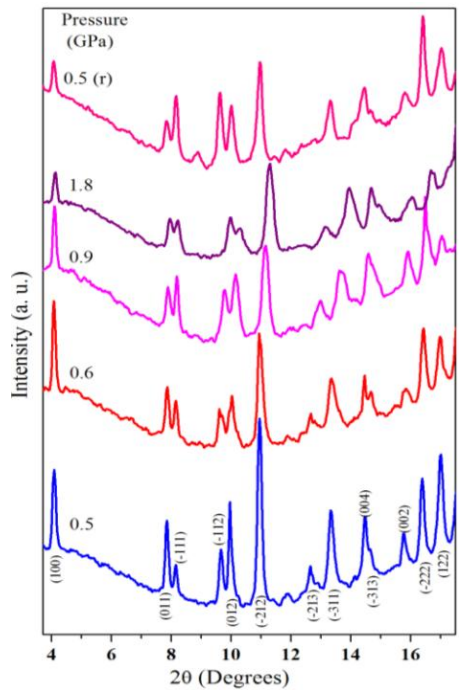

**Supplementary Figure SM1. High pressure x-ray diffraction patterns of glycinium oxalate stacked at a few representative pressures.** The XRD peaks from sample are indexed with the respective (hkl); here 'r' indicates the XRD pattern corresponding to released run.

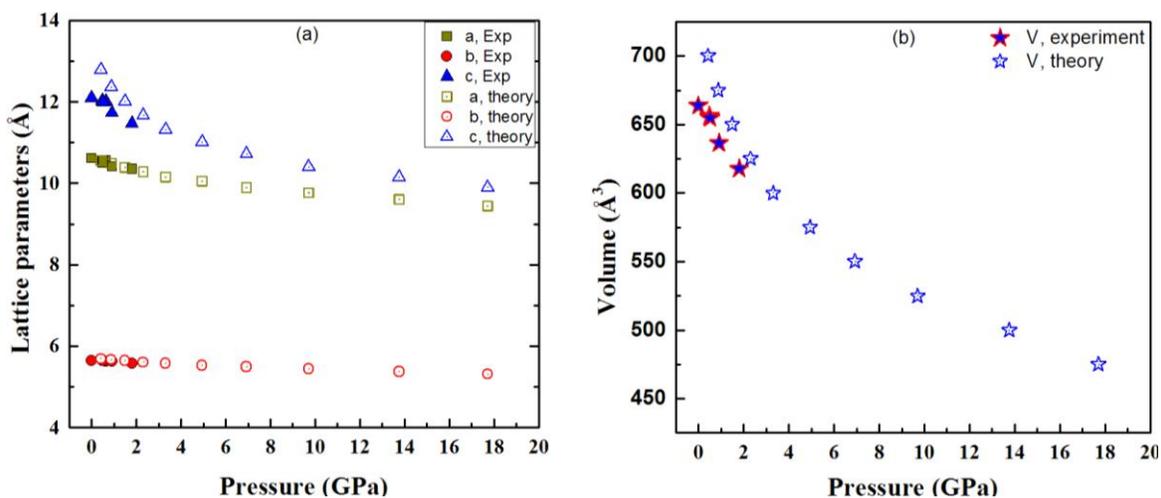

**Supplementary Figure SM2. Variation of lattice parameters (a) and unit cell volume (b) with pressure.** Filled and open symbols correspond to experiment and theory. The *a* and *c* parameters approach each other with pressure (this is accompanied with reduction in *β* angle, (~113° at ambient)).

## S6. O1-H1---O5 hydrogen bond under pressure

The calculated parameters of O1-H1---O5 hydrogen bond between semi-oxalate and glycine units and the variation of corresponding OH stretching IR mode under pressure indicate strengthening of O1-H1---O5 hydrogen bond at higher pressures (Fig. SM3).

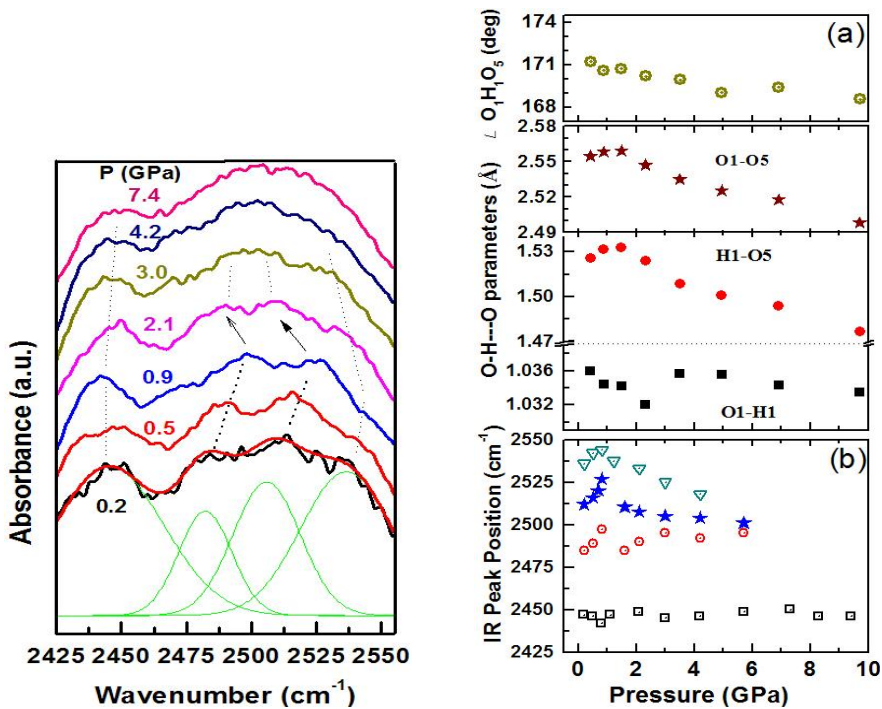

**Supplementary Figure SM3. Left: Infrared spectra of glycinium oxalate in the 2425 – 2560 cm$^{-1}$ region at high pressures.** Spectra have been offset for clarity. Arrows are guide to the eye. Deconvoluted peaks in the lowest pattern are Gaussian fits to the spectrum; **Right: (a) Calculated bond parameters of O1-H1---O5 hydrogen bond in GO at various pressures and (b) Variation of observed IR peaks with pressure in the spectral range 2425 – 2550 cm$^{-1}$.** The peak near 2510 cm$^{-1}$ may be linked to the O1-H1 stretching mode based on frequency-bond length correlation curves [23] and theory.

## S7. Planar conformation of glycine at high pressures

In the *ac*-plane, two adjacent glycine molecules are connected in a head to tail configuration (through N1-H5---O2), which are described here as bilayers (Fig. 1 of main text). At high pressures, the inter-bilayer O2(gly)---O2(gly) distance (more than 4 Å at ambient pressure) consistently decreases (Fig. SM4 inset), and reaches the limiting value ~ 3 Å at ~ 2.5 GPa, ~2.9

Å at 8 GPa and 2.8 Å (extreme limit) above 12 GPa. Under compression, to overcome the steric repulsion in the *ac*-plane, molecular reorientations would be necessitated. In the observed IR and Raman spectra, the pressure induced reduction in the separation between the $\nu_{sym}$ and $\nu_{asy}$ CH$_2$ IR and Raman modes (~60 to 40 cm$^{-1}$ upto 8 GPa) (Fig. 5b of main text) and the corresponding C-H bond distances (Fig. SM4) indeed indicate systematic changes in the conformation of glycine from bent to planar [41, 63]. This is also evident from the simulated structure at higher pressures (Fig. 1 top right of main text).

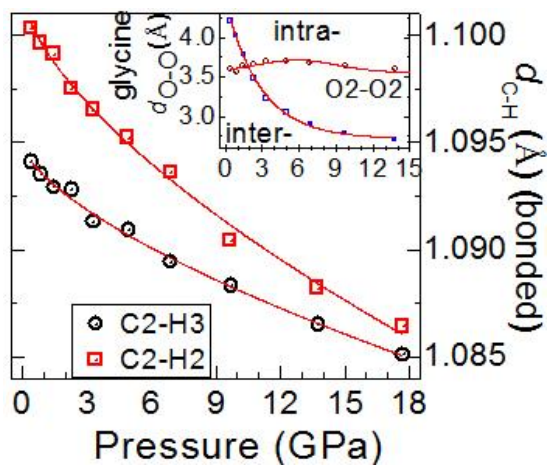

**Supplementary Figure SM4. Variation of bonded glycine CH distances and inset: non-bonded glycine O2-O2 distances (inter and intra bilayer described in Fig.1 of main text) with pressure.**

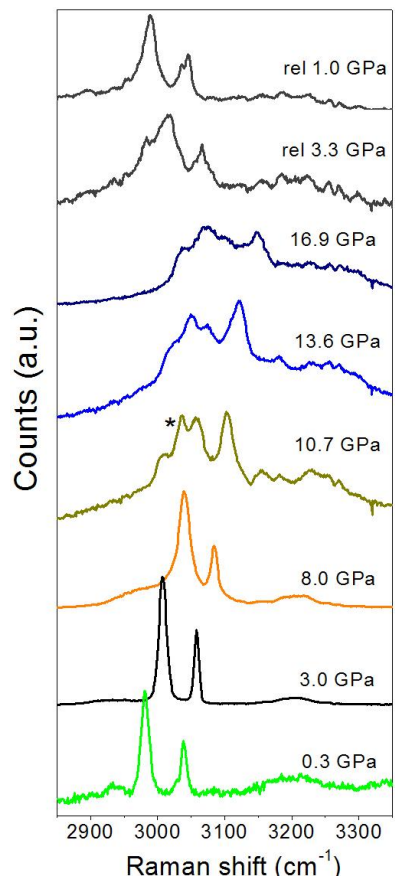

**Supplementary Figure SM5. High pressure Raman spectra of GO at some pressures in the region 2850 – 3350 cm$^{-1}$.** Here asterisk (*) denotes new features emerged at higher pressures due to change in glycine conformation [41, 63] and 'rel' denotes release pressure data. Note also the increase in background of profile in the NH stretching region above 10 GPa due to transformation towards a supramolecular phase.